\documentclass[a4paper,11pt]{article} 
\usepackage[margin=0.75in]{geometry} 
\usepackage{amssymb}
\usepackage{amsmath,epsfig, epsf, epic, graphicx} 
\usepackage{fancybox, lscape} 
\usepackage{bm}
\usepackage{natbib}
\usepackage{graphicx}
\DeclareGraphicsExtensions{.eps,.pdf,.jpeg,.png}
\usepackage{caption}
\usepackage[margin=2cm]{caption}
\usepackage{setspace}
\usepackage{xcolor}
\usepackage{array}
\usepackage{wrapfig}
\usepackage{multirow}
\usepackage{url}
\usepackage{tabularx}
\usepackage{hyperref}
\usepackage[markup=underlined, defaultcolor=black]{changes}
\usepackage{lipsum}

\allowdisplaybreaks

\usepackage{floatrow}
\usepackage{subfig}

\title{Causal Analysis of Air Pollution Mixtures: Estimands, Positivity, and Extrapolation}

\author{Joseph Antonelli\footnote{Assistant Professor, Department of Statistics, University of Florida (Email: jantonelli@ufl.edu)} and Corwin Zigler\footnote{Associate Professor, Department of Statistics and Data Science, University of Texas at Austin}}
\date{}

\begin{document}

\maketitle

\begin{abstract}
    Causal inference for air pollution mixtures is an increasingly important issue with appreciable challenges.  When the exposure is a multivariate mixture, there are many exposure contrasts that may be of nominal interest for causal effect estimation, but the complex joint mixture distribution often renders observed data extremely limited in their ability to inform estimates of many commonly-defined causal effects.  We use potential outcomes to 1) define causal effects of air pollution mixtures, 2) formalize the key assumption of mixture positivity required for estimation and 3) offer diagnostic metrics for positivity violations in the mixture setting that allow researchers to assess the extent to which data can actually support estimation of mixture effects of interest.  For settings where there is limited empirical support, we redefine causal estimands that apportion causal effects according to whether they can be directly informed by observed data versus rely entirely on model extrapolation, isolating key sources of information on the causal effect of an air pollution mixture.  The ideas are deployed to assess the ability of a national United States data set on the chemical components of ambient particulate matter air pollution to support estimation of a variety of causal mixture effects.  
\end{abstract}

\section{Introduction}
Studying the impact of air pollution on public health is but one epidemiological objective being actively enriched by the appreciation that humans are not exposed to individual contaminants in isolation, but rather to possibly complex mixtures of air pollutants \citep{dominici2010opinion, carlin2013unraveling,gibson2019overview, joubert2022powering}.  Propelled by decades of evidence establishing associations between air pollution and increased mortality and morbidity \citep{dockery1993association, pope2002lung, samet2000fine, samet2000national,portney1990policy,us2011benefits}, there is continued interest in the development of statistical methods to analyze health impacts of exposure to air pollution (and other environmental) mixtures \citep{carlin2013unraveling,gibson2019overview, joubert2022powering}. These methods set out to answer a number of questions around the effect of individual components of a mixture, interactions between mixture components, and the cumulative effect of the mixture \citep{braun2016can}.

Increased emphasis on statistical methods for complex mixtures is co-evolving with the continued adoption of explicit causal inference methods in environmental epidemiology \citep{sommer2021assessing, carone2020pursuit}. Key intersections of these methodological perspectives appear in \cite{wilson2018model}, who tailor Bayesian Model Averaging to confounding uncertainty and adjustment in an exposure wide association study, \cite{keil2020quantile}, who generalized weighted quantile sum regression \citep{carrico2015characterization} using principles from g-computation to estimate causal effects of environmental mixtures, and \cite{traini2022multipollutant}, who pursue a version of generalized propensity scores \citep{hirano2004propensity} for multipollutant causal effect estimation. While this and other work has offered important advances to causal inference with air pollution mixtures, the methods for effect estimation draw heavily from extending tools from the univariate exposure case. 

This paper takes aim at one salient challenge that is particularly pronounced in the context of air pollution mixtures, yet cannot be addressed with obvious extensions from the univariate setting. We focus on a) defining causal estimands for air pollution mixtures and b) offering a data-driven way to assess the extent to which observed data actually carry empirical support from which to estimate causal effects of interest.  Linking such considerations to policy relevance is essential, with clear definition of estimands, explicit acknowledgement and discussion of crucial assumptions, and the separation of the causal effect from a statistical model or parameter among the motivating pushes for causal inference \citep{dominici2017best, carone2020pursuit}. Convincing acknowledgment of this appears in
\cite{keil2021bayesian}, who estimate the causal effect of a reduction in the air pollution mixture hypothesized to occur if a set of power plants were to be decommissioned.  Discussion of that work in \cite{zigler2021invited} pointed towards the tension between defining estimands of policy relevance and the threat that observed data contain little or no information to actually estimate them without model-based extrapolation.  


We elaborate how the construction of mixture estimands can quickly lead to violation of the fundamental assumption of positivity, that is, that observations have positive probability of being exposed to levels of a mixture that constitute a causal estimand. We highlight the ease with which positivity violations can arise in studies of air pollution mixtures, and how such violations can produce biased effect estimates that cannot be remedied with flexible statistical models. Our proposal consists of two related efforts: data-driven diagnostics are provided to assess positivity violations with multivariate mixtures, and these diagnostics are then used to provide alternative paths forward with redefined causal estimands for mixtures. We use ambient particulate matter chemical component data across the United States to highlight how easily these difficulties can arise in practice. Throughout, we focus on air pollution mixtures, but key considerations apply for general environmental mixtures of similar dimension.

\section{Causal estimands for environmental mixtures}
\subsection{Which Mixture Effect?  Anchoring Causal Estimands to Interventions}
Throughout, we consider data observed as $(Y_i, \boldsymbol{W}_i, \boldsymbol{X}_i)$ for $i=1, \dots, n$ observations. Let $Y_i$ be a continuously scaled outcome of interest, but the key ideas hold for other outcome types. Let $\boldsymbol{W}_i$ represent a q-dimensional set of exposures that make up the air pollution mixture. In air pollution studies, $q$ is typically of moderate size, so while the methods here are general, they are most practical for $q \leq 10$. We also observe a p-dimensional vector of covariates $\boldsymbol{X}_i$ that can be used to adjust for confounding of the relationship between $\boldsymbol{W}_i$ and $Y_i$. Formulating causal estimands begins with defining $Y_i(\boldsymbol{w}_i)$ as the potential outcome that would be observed for unit $i$ had exposure for that unit been $\boldsymbol{w}_i$. Implicit in this definition of potential outcomes is the stable unit treatment value assumption (SUTVA, \cite{little2000causal}), encoding that the value of $Y_i(\boldsymbol{w}_i)$ does not depend on the manner in which the value $\boldsymbol{w}_i$ is realized (no multiple versions of treatment) and that the potential outcome for unit $i$ only depends on $\boldsymbol{w}_i$ and not exposures to other units. 

Causal estimands can be defined as any comparison between potential outcomes under two different values of $\boldsymbol{w}$.  However, the nature of multivariate exposures introduces the key question: Which differences in $\boldsymbol{w}$ are actually of interest? While earlier development focused on estimating effects of changes in one mixture component at a time, there is growing appreciation that such estimands may not represent the full implications of changes in the mixture or have any bearing on how a mixture is expected to change in response to any practicable intervention \citep{keil2021bayesian, smith2023estimating}. Throughout, we formulate causal estimands as comparisons between observed levels, $\boldsymbol{W}_i$,  and levels that might plausibly arise under some specified intervention of interest, $\boldsymbol{W}_{i, int}$. Importantly, the action needn't have actually occurred, such as in the case study of \cite{keil2021bayesian} that assumed a uniform proportional reduction in six airborne metals that was expected under the decommissioning of nearby power plants.  Alternatively, $\boldsymbol{W}_{i, int}$ may be linked to an actual change in mixture observed following an intervention, such as in \cite{nethery2021evaluation} who targeted the impact of the clean air act amendments. Using a specific intervention has the dual purpose of anchoring inference to a practicable action and isolating from the infinitely many hypothetical changes in $\boldsymbol{W}$. An additional benefit, which we prove important in subsequent discussion, is that this helps to ensure that interventions correspond to mixture levels that do not stray too far from those naturally observed. Note that there may be other scientific questions of interest in the analysis of mixtures, such as those studied in \cite{gibson2019overview}, that may not require formalization as effects of interventions.

Let $\boldsymbol{\Delta}_i = \boldsymbol{W}_i - \boldsymbol{W}_{i,int}$ denote the change in the environmental mixture for unit $i$ in response to the intervention. Note that this shift is unit specific to accommodate the case where an intervention changes the mixture differently across observations. For instance, an intervention to reduce pollution emissions from a point source may differentially impact locations at various distances from the source. The causal estimand defining the effect of interest can be expressed as: \begin{equation}\frac{1}{n} \sum_{i=1}^n \bigg\{ Y_i(\boldsymbol{W}_i) - Y_i(\boldsymbol{W}_{i, int}) \bigg\}, \label{exp:estimand}
\end{equation} which represents the average impact of the change in mixture following the specified action across the $n$ observations under study. While we focus on (\ref{exp:estimand}), the considerations below apply to any estimand comprised of a pre-specified contrast in $\boldsymbol{W}$. 

\subsection{Key assumptions and the importance of mixture positivity}
Estimating (\ref{exp:estimand}) with observed data critically rests on several assumptions.  Chief among them in observational studies is the assumption of no unmeasured confounding, which states that there are no unmeasured common causes of the mixture and the outcome. The importance of this assumption means that it must be carefully evaluated within the context of any study.  Despite this importance, strategies for confounding adjustment are not the focus of this work, in part because confounding considerations are not unique to studies of environmental mixtures, with ample methodologies available from the context of a single exposure.

The final foundational assumption required for the causal inference - and the one on which we focus most here - is that of {\it positivity}. Letting $f(\boldsymbol{w} \vert \boldsymbol{X} = \boldsymbol{x})$ denote the density of the exposures conditional on the covariates taking value $\boldsymbol{x}$, positivity in our setting can be defined as
$$f(\boldsymbol{W}_{i,int} \vert \boldsymbol{X} = \boldsymbol{X}_i) > 0 \quad \text{for all } \boldsymbol{W}_{i,int}, \boldsymbol{X}_i.$$
Note that our positivity assumption is unique to the sample level estimand in \ref{exp:estimand}, and other estimands would require a modified assumption. Many of the ideas presented here, however, extend to other notions of positivity. Positivity violations can be categorized into either structural positivity violations that occur when certain exposure values are not possible for certain covariate values, or finite sample positivity violations that come from not observing certain exposures for a particular covariate value in the sample, even if these are hypothetically possible \citep{petersen2012diagnosing}. We focus on the latter of these two throughout as it is more common in air pollution epidemiology. The essence of the positivity assumption states that the observed data contain empirical support for estimating values of $Y_i(\boldsymbol{W}_{i,int})$ for all units: inferring the unobserved potential outcome under $\boldsymbol{W}_{i,int}$ for unit $i$ requires observations with similar covariate values having been observed with that value of the mixture. Violations of positivity correspond to the absence of such information, in which case inference for causal effects must rely on extrapolation, typically using a parametric model.  This assumption introduces threats to causal validity for air pollution mixtures that can be more salient than in studies of univariate exposures.

\section{Illustrating Positivity Violations and Model Extrapolation for Environmental Mixtures}
\label{sec:Danger}

\begin{figure}[t]
    \centering
    \includegraphics[width=0.85\linewidth]{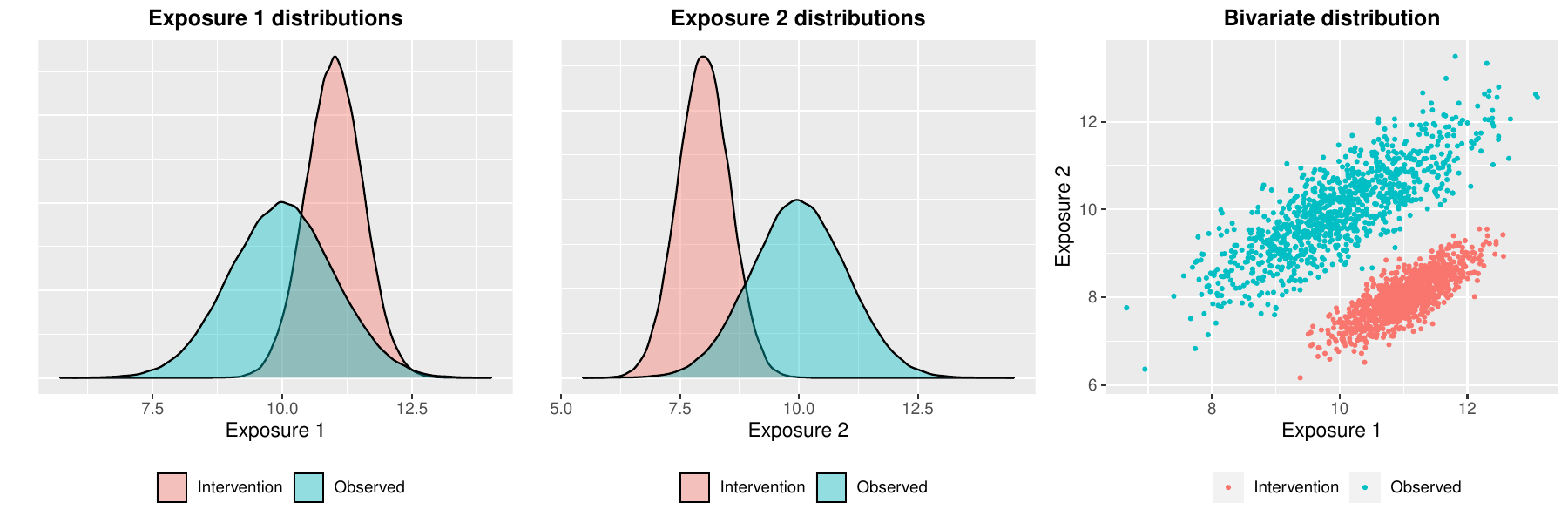}
    \caption{The left panel shows the distribution of exposure 1 in the observed data and in the intervention distribution. The middle panel shows the same distributions, but for exposure 2. The right panel shows the bivariate distribution for exposures 1 and 2.}
    \label{fig:BivariateOverlap}
\end{figure}

In the case of a univariate exposure, there is ample literature on the positivity assumption and the threats to validity that can arise amid its violations \citep{crump2009dealing, lee2011weight, li2018balancing}. To illustrate how these issues are exacerbated amid multidimensional consideration of whether a value of $\boldsymbol{W}_{int}$ lies within the observed joint distribution of a multivariate mixture, we simulate data with no covariates, $q=2$ exposures, and a moderately nonlinear exposure-response curve. We use a large data set of $n=1000000$ to minimize sampling variability. Figure \ref{fig:BivariateOverlap} shows both the marginal distributions of the two exposures, as well as the joint distribution of the two exposures for both the observed data and a hypothetical intervention distribution. Judging only from the marginal distributions, which would be analogous to positivity assessment in the univariate case, there is substantial overlap in the observed and intervention distribution for exposure 1, and a moderate degree of overlap for exposure 2. Thus, marginal investigation of each mixture component indicates that values of $\boldsymbol{W}_{i,int}$ lie within the observed data. However, investigating the joint mixture distribution clearly indicates otherwise, with values of $\boldsymbol{W}_{i,int}$ falling completely outside of the observed distribution. 

To illustrate the consequences of this lack of mixture overlap, we estimate the treatment effect using both a linear model and nonlinear models with 3 and 5 degrees of freedom natural cubic splines, each allowing for interactions between the mixture components. For full details of the estimation strategies and data generating mechanism for this simulation study, see Appendix A. The true causal effect defined by \ref{exp:estimand} in this scenario is 1.45, the linear model estimates it to be 1.22, the 3 degree of freedom model estimates it to be 1.17, and the 5 degree of freedom model estimates it to be -0.37. The lack of overlap leads to extrapolation, which amplifies bias due to model misspecification, and we obtain poor estimates of the causal effect, with the more flexible models performing worse. This example highlights a number of relevant issues for environmental mixtures. First, notions of overlap borrowed from the univariate exposure case do not imply overlap in the mixture. Second, overlap for mixtures with $q > 2$, will be increasingly difficult to visualize and diagnose. Lastly, although we have omitted them for illustration, overlap and positivity must be evaluated conditional on covariates, $\boldsymbol{X}$, which will be even more restrictive and less likely to hold in practice.

\section{Potential paths forward: Diagnosing Mixture Positivity and Alternative Estimands}\label{sec:estimands}

\begin{figure}[t]
    \centering
    \includegraphics[width=0.55\linewidth]{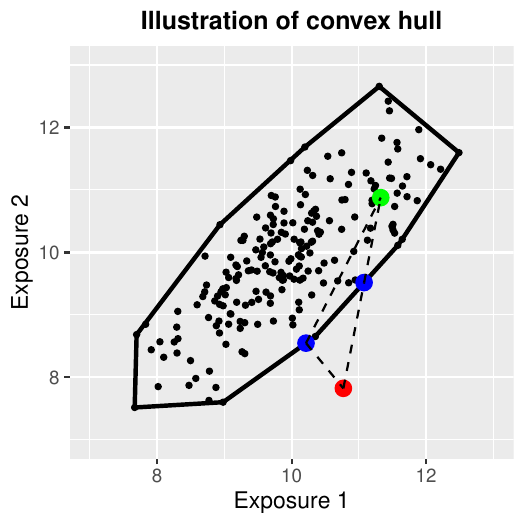}
    \caption{Illustration of the convex hull on a subset of the observed exposures from Section 3. The green point is an observed exposure of interest, $\boldsymbol{W}_i$. The red point is the corresponding interventional point of interest, $\boldsymbol{W}_{i,int}$. The blue points represent two different feasible values of the exposure that fall in the convex hull as described in Section \ref{sec:FeasibleSub}.}
    \label{fig:ConvexHullToy}
\end{figure}

The preceding section motivates the need for formal diagnostics to identify and address problems due to positivity violations for environmental mixtures. We focus on situations without covariates $\boldsymbol{X}$, though we emphasize that these same issues are likely exacerbated in the presence of confounders and discuss extensions in Section \ref{sec:Discussion}. Following similar ideas to diagnose model extrapolation presented in \cite{king2006dangers}, we utilize the concept of the convex hull of multivariate exposures. Intuitively, the convex hull of the exposures is the smallest polygon that contains all of the observed exposure values and therefore it can be thought of as the region where exposure values are actually observed. See Figure \ref{fig:ConvexHullToy} for an illustration in the case of $q=2$. If $\boldsymbol{W}_{i,int}$ lies outside of the convex hull, then extrapolation is necessarily required to estimate potential outcomes under that value of  $\boldsymbol{W}_{i,int}$.

For each observation, let $\boldsymbol{W}_{i,hull}$ denote the point in the convex hull that is closest to the intervention point $\boldsymbol{W}_{i,int}$. Then, a metric quantifying how far the intervention point lies from the convex hull can be calculated as:
$$R_i = \frac{\text{distance}(\boldsymbol{W}_{i,hull}, \boldsymbol{W}_{i,int})}{\text{distance}(\boldsymbol{W}_i, \boldsymbol{W}_{i,int})},$$
where $\text{distance}(\boldsymbol{w}_1, \boldsymbol{w}_2)$ is the euclidean distance between points $\boldsymbol{w}_1$ and $\boldsymbol{w}_2$. The value of $R_i$ is necessarily between 0 and 1, with larger values implying a heavier reliance on model extrapolation, because the distance from the convex hull is large. The best value, $R_i=0$, implies no model extrapolation because the intervention point already lies in the convex hull. The distribution of $R_i$ across the sample can indicate the extent to which estimates of the quantity in (\ref{exp:estimand}) rely on model extrapolation. If $R_i$ is small, such as less than 0.1, for all observations, then causal estimates will not be greatly impacted by extrapolation. In contrast, high values of $R_i$ in the sample indicate a threat of extrapolation and motivate alternative estimands such as those proposed in the following sections.

\subsection{Alternative estimand one: Effects of feasible exposure changes}
\label{sec:FeasibleSub}
Upon diagnosing that estimation of (\ref{exp:estimand}) would rely on model extrapolation, alternatively-defined estimands defined based on the convex hull may be of interest. One such estimand is to find a ``feasible'' level of exposure, $\boldsymbol{W}_{i, feas}$, that is similar to the intervention value of interest but not subject to model extrapolation. We explore two options for feasible values:
\begin{enumerate}
    \item $\boldsymbol{W}_{i, feas} = \boldsymbol{W}_{i,hull}$ so that the feasible value is the point in the convex hull that is closest to $\boldsymbol{W}_{i,int}$
    \item Find the largest value of $\phi$ between 0 and 1 such that $\boldsymbol{W}_{i, feas} = \boldsymbol{W}_{i} + \phi (\boldsymbol{W}_{i, int} - \boldsymbol{W}_i)$ is in the convex hull. 
\end{enumerate}
An illustration of these two values can be found in Figure \ref{fig:ConvexHullToy}. The first specifies a mixture within the hull of the observed mixture value that is closest to the hypothesized mixture under the intervention. The second value represents a shift in the exposure mixture in the same direction as the hypothesized value under the intervention, but smaller in magnitude to remain within the hull of the observed data. In either case, the estimand in (\ref{exp:estimand}) can be decomposed as:
\begin{align*}
    \frac{1}{n} \sum_{i=1}^n \bigg\{ Y_i(\boldsymbol{W}_i) - Y_i(\boldsymbol{W}_{i, int}) \bigg\} &= \underbrace{\frac{1}{n} \sum_{i=1}^n \bigg\{ Y_i(\boldsymbol{W}_i) - Y_i(\boldsymbol{W}_{i, feas}) \bigg\}}_{\text{feasible estimand}} + \underbrace{\frac{1}{n} \sum_{i=1}^n \bigg\{ Y_i(\boldsymbol{W}_{i, feas}) - Y_i(\boldsymbol{W}_{i, int}) \bigg\}}_{\text{extrapolation component}}.
\end{align*}
This decomposition explicitly delineates the part of the causal effect that can be estimated with empirical support from the observed data from the part that is only available through extrapolation.

Table \ref{tab:Estimands} displays estimates of these effect components using the simulated data from Section \ref{sec:Danger}. We see that the linear model underestimates both the overall effect and the feasible component, which is expected owing to the nonlinearity of the simulated outcomes. The nonlinear models provide estimates closer to the true value of the feasible component, but get increasingly worse at estimating the extrapolation component as the complexity of the model grows. This illustration highlights the importance of decomposing (\ref{exp:estimand}); The feasible estimand is clearly less sensitive to model specification, while the extrapolation component is highly sensitive, with different models providing estimates very far from the truth as misspecification bias becomes amplified in the presence of extrapolation. The median value of $R_i$ in these data is 0.34, suggesting a high degree of model extrapolation that ultimately yields sensitivity to model choice. 

\begin{table}[ht]
\centering
\begin{tabular}{rrrrr}
  \hline
 & True effect & linear model & 3 df model & 5 df model \\ 
  \hline
Overall effect & 1.45 & 1.22 & 1.17 & -0.37 \\ 
  Feasible estimand & 0.81 & 0.68 & 0.75 & 0.74 \\ 
  Extrapolation component & 0.64 & 0.54 & 0.42 & -1.11 \\ 
   \hline
\end{tabular}
\caption{Estimates of treatment effects under different statistical models from the simulated data set of Section \ref{sec:Danger} when looking at feasible estimands.}
\label{tab:Estimands}
\end{table}

\subsection{Alternative estimand two: Effects in feasible subpopulations}

An alternative option is to focus on a subpopulation with feasible values of $\boldsymbol{W}_{i, int}$. We can define estimands as:
$$\sum_{i=1}^n \gamma_i \bigg\{ Y_i(\boldsymbol{W}_i) - Y_i(\boldsymbol{W}_{i, int}) \bigg\},$$
where the $\gamma_i$ are positive and sum to 1, i.e. $\sum_{i=1}^n \gamma_i = 1$. The  effect in (\ref{exp:estimand}) is a special case where $\gamma_i = 1/n$ for all $i$, implying that all data points are assigned equal weight. However, when confronting lack of overlap, weights can be incorporated so that observations requiring more extrapolation receive less weight in the definition of the causal effect. One option is to specify weights $\gamma_i = 0$ for all units with $R_i$ above some threshold, excluding units from the treatment effect that exhibit exposure values requiring large amounts of extrapolation. This is commonly done for binary treatments, where units with extreme propensity score estimates are trimmed from the sample before estimating treatment effects \citep{ho2007matching, cole2008constructing, huber2013performance, xiao2013comparison, kilpatrick2013exploring}. A continuous alternative would assign values of $\gamma_i$ inversely proportional to the amount of extrapolation required for each data point. For example, setting $$\gamma_i = \frac{(1 - R_i)}{\sum_{j=1}^n (1 - R_j)},$$
would assign the largest weights to points requiring no extrapolation, and decreasing weight as the amount of extrapolation increases. These weighted estimands present a trade-off between interpretability and bias with respect to the overall effect in (\ref{exp:estimand}). While less susceptible to bias amplified by model extrapolation, 
they estimate an effect in some weighted subpopulation that may be difficult to understand or describe. Table \ref{tab:WeightedEstimands} shows estimates from the simulated data from Section \ref{sec:Danger} when using $\gamma_i = 1/n$ and when using weights that are trimmed so that only observations with $R_i < 0.05$ are included. We see that the true value of the estimand changes, but importantly, the lack of required extrapolation leads all models to provide similar estimates of this alternatively-defined effect. 
\begin{table}[ht]
\centering
\begin{tabular}{rrrrr}
  \hline
 & True effect & linear model & 3 df model & 5 df model \\ 
  \hline
Equal weights & 1.45 & 1.22 & 1.17 & -0.37 \\ 
  Trimmed weights & 0.79 & 0.66 & 0.73 & 0.71 \\ 
   \hline
\end{tabular}
\caption{Estimates of treatment effects under different statistical models from the simulated data set of Section \ref{sec:Danger} when looking at trimmed estimands.}
\label{tab:WeightedEstimands}
\end{table}

\section{Illustration on United States air pollution mixtures}
We illustrate the ideas above in the context of estimating causal effects of ambient PM$_{2.5}$ component mixtures. We only consider exposure data, as the approach would hold direct relevance estimating the impact of these mixture components on any health outcome. Extensions to adjust for observed confounding are discussed in Section \ref{sec:Discussion}. 

Air pollution exposure data for the year 2015 are obtained from the Atmospheric Composition Analysis Group \citep{van2019regional}. We utilize data on annual average levels of black carbon (BC), organic matter (OM), ammonium (NH4), nitrates (NIT), and sulfates (SO4) within the contiguous United States. We examine commonly-specified causal mixture effects with values of $\boldsymbol{W}_{i,int}$ corresponding to a) reducing a single mixture component by a specified proportion, or b) simultaneously reducing all mixture components by a specified proportion. In both cases, the proportion specified is varied from 10\% - 90\% to investigate increasingly pronounced reductions and utilize the metrics defined in previous sections to explore degree of empirical support for estimating the health effects of such shifts in exposures. Note that while we only focus here on settings where all components of the mixture are shifted in the same direction, the same ideas would hold for any shift in exposures caused by the intervention.

\begin{figure}[t]
    \centering
    \includegraphics[width=0.6\linewidth]{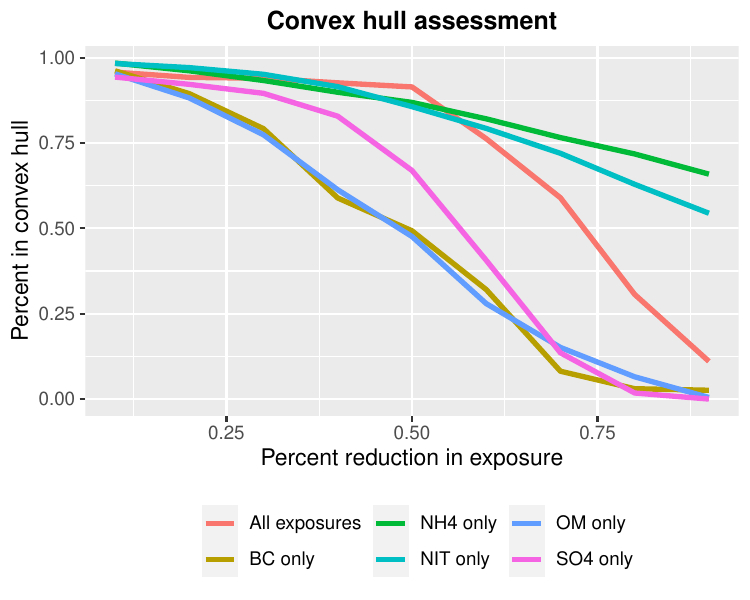}
    \caption{Percentage of $\boldsymbol{W}_{i, int}$ values that fall within the convex hull of the observed mixture data.}
    \label{fig:PercentInHull}
\end{figure}

Figure \ref{fig:PercentInHull} shows the percentage of $\boldsymbol{W}_{i, int}$ values that fall inside of the convex hull of the observed exposure data as the exposure reduction becomes more pronounced. As expected, more extreme reductions result in fewer values of $\boldsymbol{W}_{i,int}$ falling within the convex hull. The degree to which empirical support suffers varies according to the specific estimand: specifying a reduction in only OM or BC yields many values of $\boldsymbol{W}_{i,int}$ outside the convex hull even under relatively small reductions. Other components, such as NIT or NH4, could be specified with large exposure reductions and still contain a majority of implied values of $\boldsymbol{W}_{i,int}$ within the convex hull.  The estimand specifying simultaneous reduction of all mixture components maintains nearly all observations within the convex hull up to percent reductions of 50\%, but larger reductions lose empirical support quickly. 

\begin{figure}[t]
    \centering
    \includegraphics[width=0.8\linewidth]{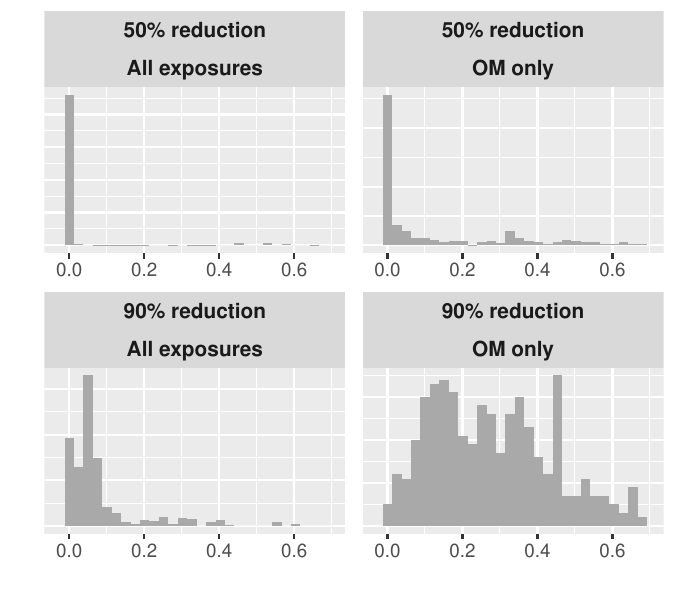}
    \caption{Distribution of $R_i$ when targeting an intervention that reduces exposure to either organic matter only, or all components of the mixture simultaneously.}
    \label{fig:RdistributionPM}
\end{figure}

An examination based on distance from the hull encoded by $R_i$ is depicted in Figure \ref{fig:RdistributionPM}.  Results echo those of Figure \ref{fig:PercentInHull}; A 50\% reduction in all components shows most values of $R_i$ at or near 0, but assuming a 90\% reduction yields more values of $R_i$ further from zero. The estimand considering a 50\% reduction in only OM has many values of $R_i$ far from zero, with a 90\% reduction in OM corresponding to an extreme case where extrapolation is required to estimate the effect of such a reduction.  

This illustration presents three key points related to estimating causal effects of PM$_{2.5}$ component mixtures.  First, the degree to which a particular mixture estimand can be supported with empirical observations vs. model extrapolation is dependent on which component(s) are of interest.  Second, larger reductions in exposure will be harder to estimate without extrapolation, with most estimands exhibiting very little empirical support for reductions greater than 50\%.  Finally, note that the simultaneous estimand indicates more robustness to model extrapolation than some of the estimands corresponding to single component reductions, highlighting the importance of considering the entire joint mixture distribution when diagnosing positivity violations and empirical support for causal effect estimation. Ultimately, the usefulness of the alternatively-defined estimands in Section \ref{sec:estimands} must be judged in light of the degree of extrapolation and the implied differences with the overall effect in (\ref{exp:estimand}).

\section{Discussion}
\label{sec:Discussion}

We have emphasized the importance of defining relevant causal estimands for air pollution mixtures and formalized how positivity violations and reliance on model extrapolation can challenge estimation. Relative to other important methodological considerations for causal inference that can follow development in the case of a univariate exposure, these issues present a particularly pressing challenge when attempting causal inference with mixtures. 

Using the convex hull of observed exposures, we quantified the extent to which a specified mixture estimand can be supported by available data. This work has conceptual links to previous work assessing the dangers of model extrapolation when drawing causal inferences across regions of covariate space with limited representation for each of two treatment groups \citep{king2006dangers}. In the context of air pollution mixtures, when estimands of initial interest correspond to mixture values that are not represented in the observed data, we propose novel alternative estimands as useful quantities for understanding the impact of environmental mixtures that are less reliant on model extrapolation.  Even when some degree of extrapolation is necessary - which is a common setting in the analysis of mixtures - it is essential to assess sensitivity to model specification and evaluate the degree to which ultimate inferences rely on modeling assumptions vs. empirical support from the observed data. Additionally, researchers must thoroughly consider the interpretability trade-off with estimands that shift focus away from the causal effect of original interest to ensure that alternative estimands less susceptible to extrapolation and misspecification retain public health relevance.

The theoretical underpinnings of our discussion of air pollution mixtures extend to more general forms of environmental mixtures.  The dominant practical consideration relates to the number of mixture components, which could far exceed that typical to air pollution mixtures in, for example, exposome-wide association studies. While the diagnostics proposed in this manuscript could still be performed, they will likely indicate very little information in the data to estimate causal contrasts of high-dimensional exposures, and the proposed alternative estimands may be of limited use. Additionally, the present work focuses on positivity violations that require extrapolation outside the convex hull of the exposure data, but violations can also arise within the convex hull when exposure values in the hull still differ from observed values. This can give rise to the need for {\it interpolation} to support estimation of causal effects, which, while distinct from extrapolation, might prove equally salient in high-dimensional settings where sparsity within the convex hull of exposures persists.  Additional metrics to quantity the degree of sparsity and interpolation are warranted.
For simplicity, the exposition here did not explicitly consider simultaneous confounding adjustment, which would always be required in observational settings.  In technical terms, our discussion focused on marginal positivity with respect to the marginal exposure distribution, whereas notions of conditional positivity (conditional on covariate values,  $\boldsymbol{X}$) are analogous and essential. Conditional positivity represents a more restrictive assumption, and therefore any issues with model extrapolation identified in marginal distributions of the exposures are expected to be exacerbated when additionally conditioning on covariates. The conceptual extension to confounding adjustment and conditional positivity is straightforward, with calculations of the convex hull available to be carried out within strata of observed confounders. In practice, the calculation of the convex hull could be extended to include the hull of $\boldsymbol{W}$ and some or all components of a covariate vector, $\boldsymbol{X}$.  Or, covariate summaries such as those based on the generalized propensity score for multivariate exposures could be included in the convex hull calculation.  A further investigation of these extensions is a topic for future research. Nonetheless, we hope that the discussion of the marginal case can help focus attention on positivity and extrapolation with exposure mixtures and that the provided tools can anchor continued progress in causal effect estimation in the context of environmental mixtures.

\section*{Acknowledgements}

Research described in this article was conducted under contract to the Health Effects Institute (HEI), an organization jointly funded by the United States Environmental Protection Agency (EPA) (Assistance Award No. CR-83590201) and certain motor vehicle and engine manufacturers. The contents of this article do not necessarily reflect the views of HEI, or its sponsors, nor do they necessarily reflect the views and policies of the EPA or motor vehicle and engine manufacturers. 

\bibliographystyle{apalike}
\bibliography{refs.bib}

\appendix

\section{Details of Simulation Study}

\begin{figure}[t]
    \centering
    \includegraphics[width=0.55\linewidth]{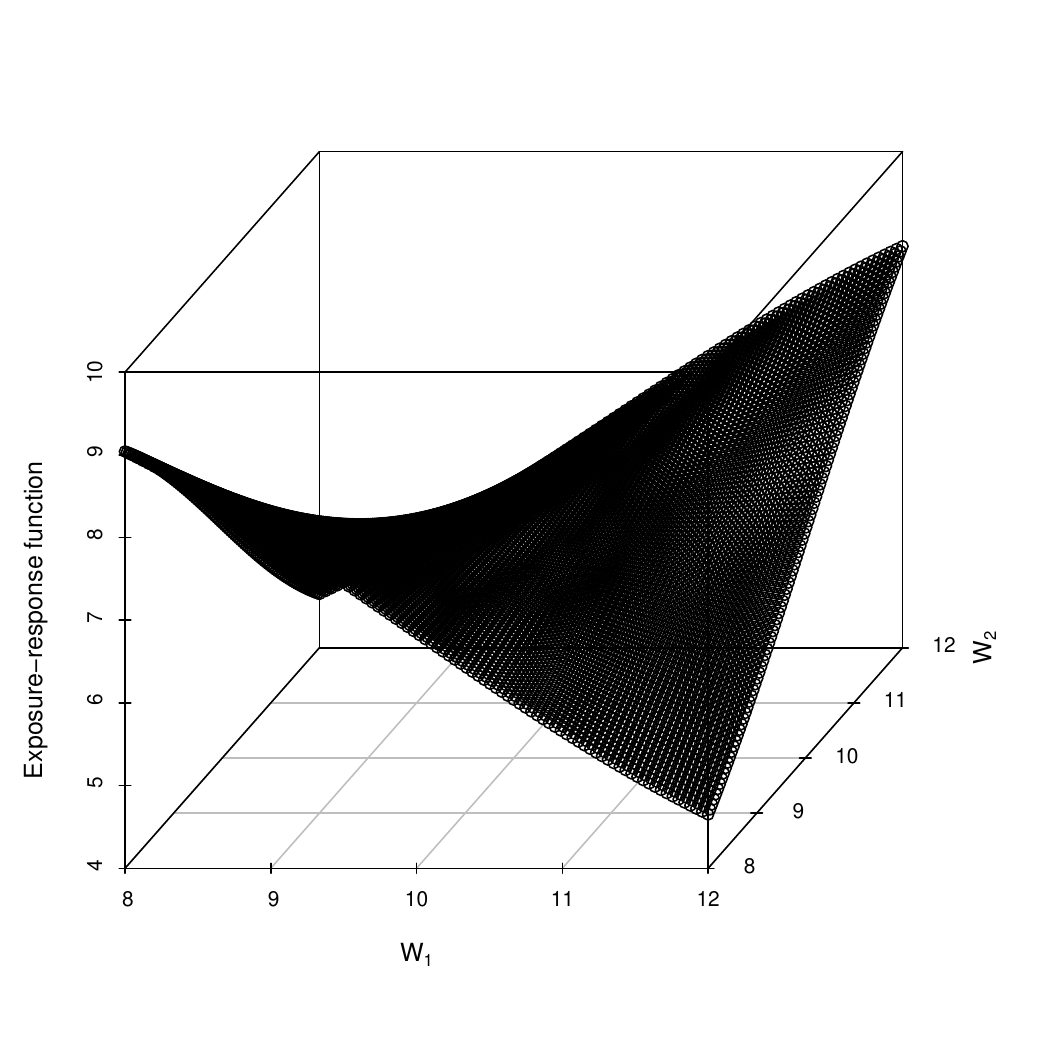}
    \caption{Illustration of the $g(\boldsymbol{w})$ function.}
    \label{fig:BivariateERcurve}
\end{figure}

Here we provide more details on the data generating mechanism utilized in the simulation study in Section 3 of the manuscript, as well as details on the different estimation strategies used to estimate causal effects. We first need to generate exposures under both the observed data and intervention distributions. We simulate exposure values using a bivariate normal distribution given by
\begin{align*}
    \boldsymbol{W}_i &\sim \mathcal{N}(\boldsymbol{\mu}, \boldsymbol{\Sigma}) \\
    \boldsymbol{W}_{i,int} &\sim \mathcal{N}(\boldsymbol{\mu}_{int}, \boldsymbol{\Sigma}_{int}).
\end{align*}
We set the means of these distributions to be $\boldsymbol{\mu} = (10, 10)$ and $\boldsymbol{\mu}_{int} = (11, 8)$. The covariance matrix for the observed data exposures is given by
\begin{equation*}
\boldsymbol{\Sigma} = 
\begin{pmatrix}
1 & 0.8 \\
0.8 & 1
\end{pmatrix},
\end{equation*}
which implies a correlation of 0.8 between the two exposures. We use the same correlation structure for the intervention exposures, but also reduce their variability, by setting $\boldsymbol{\Sigma}_{int} = 0.3 \boldsymbol{\Sigma}$. The potential outcomes are then generated by 
$$Y_i(\boldsymbol{w}) = g(\boldsymbol{w}) + \epsilon_i, \quad \epsilon_i \sim \mathcal{N}(0, 1).$$
The true causal effect in this scenario is given by
\begin{align*}
    \frac{1}{n} \sum_{i=1}^n \bigg\{ Y_i(\boldsymbol{W}_i) - Y_i(\boldsymbol{W}_{i, int}) \bigg\}
    = \frac{1}{n} \sum_{i=1}^n \bigg\{ g(\boldsymbol{W}_i) - g(\boldsymbol{W}_{i, int}) \bigg\}.
\end{align*}
We define the true $g(\boldsymbol{w})$ function to be given by the following equation:
\begin{align*}
    g(\boldsymbol{w}) = 3 - 0.2 \ \text{expit}(0.5(w_1-10)) - 0.2 \ \text{expit}(0.5 (w_2-10)) -
    8 \ \text{expit}(0.3 (w_1 - 10)(w_2 - 10)).
\end{align*}
This function is visualized in Figure \ref{fig:BivariateERcurve}, which highlights that the true relationship between the exposures and outcome is both nonlinear and includes interactions. Lastly, we can detail our estimation strategy for estimating the causal effect in this scenario. We use basis function expansions to estimate the unknown $g(\boldsymbol{w})$ function. Specifically, we fit models of the form:
\begin{align*}
    g(\boldsymbol{w}) = \alpha + \sum_{k=1}^K \beta_k \phi_k(w_1) + \sum_{k=1}^K \xi_k \psi_k(w_2) + \sum_{j=1}^K \sum_{k=1}^K \zeta_{jk} \phi_j(w_1) \psi_k(w_2).
\end{align*}
When estimating treatment effects with a linear model, we have that $K=1$, $\phi_1(w_1) = w_1$, and $\psi_1(w_2) = w_2$. When using nonlinear models, we first vary the number of basis functions $K \in \{3,5\}$, and we utilize natural cubic spline basis functions $\phi_j(w_1)$ and $\psi_j(w_2)$ for $j=1, \dots, K$ with knot locations chosen to be equally spaced quantiles of $w_1$ and $w_2$, respectively. Once we have estimated this function by estimating the unknown coefficients, our estimate of the treatment effect is simply given by
$$\frac{1}{n} \sum_{i=1}^n \bigg\{ \widehat{g}(\boldsymbol{W}_i) - \widehat{g}(\boldsymbol{W}_{i, int}) \bigg\}.$$
Analogous calculations can be done to estimate the weighted and feasible estimands. 

\end{document}